\documentclass[english,aip,reprint]{revtex4-1}
\usepackage{mathptmx}
\usepackage[LGR,T1]{fontenc}
\usepackage[latin9]{inputenc}
\usepackage{color}
\usepackage{babel}
\usepackage{verbatim}
\usepackage{textcomp}
\usepackage{amstext}
\usepackage{graphicx}
\usepackage{subscript}
\usepackage[unicode=true,pdfusetitle,
 bookmarks=true,bookmarksnumbered=false,bookmarksopen=false,
 breaklinks=false,pdfborder={0 0 1},backref=false,colorlinks=true]
 {hyperref}
\hypersetup{
 urlcolor=blue,linkcolor=red,citecolor=blue}

\makeatletter

\DeclareRobustCommand{\greektext}{%
  \fontencoding{LGR}\selectfont\def\encodingdefault{LGR}}
\DeclareRobustCommand{\textgreek}[1]{\leavevmode{\greektext #1}}
\ProvideTextCommand{\~}{LGR}[1]{\char126#1}

\makeatother

\begin{document}
\title{\textcolor{black}{The electrical conductivity of cubic (In}\textsubscript{\textcolor{black}{1-}\textit{\textcolor{black}{x}}}\textcolor{black}{Ga}\textsubscript{\textit{\textcolor{black}{x}}}\textcolor{black}{)}\textsubscript{\textcolor{black}{2}}\textcolor{black}{O}\textsubscript{\textcolor{black}{3}}\textcolor{black}{{}
films ($x\le0.18$): Native point defects, Sn-doping, and the surface
electron accumulation layer}}
\author{Alexandra Papadogianni}
\affiliation{Paul-Drude-Institut für Festkörperelektronik, Leibniz-Institut im
Forschungsverbund Berlin e.V., Hausvogteiplatz 5\textendash 7, D\textendash 10117
Berlin, Germany}
\author{\textcolor{black}{Takahiro Nagata}}
\affiliation{\textcolor{black}{National Institute for Materials Science, 1-1 Namiki
Tsukuba, 305-0044 Ibaraki, Japan}}
\author{Oliver Bierwagen}
\affiliation{Paul-Drude-Institut für Festkörperelektronik, Leibniz-Institut im
Forschungsverbund Berlin e.V., Hausvogteiplatz 5\textendash 7, D\textendash 10117
Berlin, Germany}
\date{\today}
\begin{abstract}
The alloying of the group-III transparent semiconducting sesquioxides
In\textsubscript{2}O\textsubscript{3} and Ga\textsubscript{2}O\textsubscript{3}
can lead to a modulation of the properties of the parent compounds,
e.g., the shallow- and deep-donor character of the oxygen vacancy
or the presence and absence of a surface electron accumulation layer,
respectively. In this work, we investigate the effect of alloying
on the electron transport properties of unintentionally-doped single-crystalline
and textured bixbyite (In\textsubscript{1-\textit{x}}Ga\textsubscript{\textsl{x}})\textsubscript{2}O\textsubscript{3}
thin films annealed in oxygen and vacuum with Ga contents up to $x=0.18$.
Hall effect measurements demonstrate a surprising increase in electron
density due to native defects with added Ga. This increase may be
related to the incorporation of Ga-interstitials or oxygen vacancies
induced by Ga-related unit-cell distortions. A combined investigation
based on hard and soft x-ray photoelectron spectroscopy measurements
demonstrates the existence of the surface electron accumulation layer
for all alloy films and, hence, no depletion up to $x=0.18$. Finally,
we additionally demonstrate a single-crystalline (In\textsubscript{0.92}Ga\textsubscript{0.08})\textsubscript{2}O\textsubscript{3}:Sn
film, as a possible transparent conductive oxide with a wider band
gap than that of (Sn-doped) In\textsubscript{2}O\textsubscript{3}.
\end{abstract}
\maketitle

\section{Introduction}

In\textsubscript{2}O\textsubscript{3} and Ga\textsubscript{2}O\textsubscript{3}
are transparent semiconducting materials widely studied for implementation
in devices.

On the one hand, In\textsubscript{2}O\textsubscript{3} is typically
cubic bixbyite\citep{Karazhanov_2007} and has an optically forbidden
direct band gap found to be 2.7\textendash 2.9~eV, with strong optical
absorption occurring from valence bands nearly 1~eV below the valence
band maximum (VBM).\citep{Walsh_PRL.100.167402,King_PRB.79.205211,Irmscher_10.1002/pssa.201330184}
This property renders In\textsubscript{2}O\textsubscript{3} transparent
in the visible range of the electromagnetic spectrum and is\textemdash remarkably\textemdash combined
with high electrical conductivity. In\textsubscript{2}O\textsubscript{3}
exhibits inherent \textit{n}-type conductivity, which is commonly
referred to as unintentional doping (UID) and can be significantly
enhanced by doping. The unintentional doping of In\textsubscript{2}O\textsubscript{3}\textemdash with
electron concentrations ranging from $10^{17}$ to $10^{19}\,\mathrm{cm^{-3}}$\textemdash is
typically attributed to hydrogen impurities, such as singly-charged
hydrogen interstitials, $\mathrm{H}_{i}^{+}$, and substitutional
hydrogen at an oxygen site, $\mathrm{H_{O}^{+}}$.\citep{Limpijumnong_PhysRevB.80.193202}
The origin of the unintentional conductivity of In\textsubscript{2}O\textsubscript{3},
has been further associated with oxygen vacancies, $V_{\mathrm{O}}^{2+}$,
acting as doubly-ionized shallow donors. It is still, however, debated
whether the $V_{\mathrm{O}}$ are indeed shallow donors\citep{Agoston_PhysRevB.84.045311,Buckeridge_PhysRevMaterials.2.054604,Chatratin_2019}
or deep donors that do not contribute free electrons.\citep{Lany_PhysRevLett.98.045501,Limpijumnong_PhysRevB.80.193202}
The shallow donor case is also supported by experimental results showing
that annealing in oxygen, and thereby removing the oxygen vacancies,
decreases the electron concentration in the bulk of In\textsubscript{2}O\textsubscript{3}
, whereas annealing the material in vacuum does not only restore the
electron concentration, but it further increases it.\citep{Bierwagen_doi:10.1063/1.4751854}
Indium interstitials, that are predicted to act as shallow donors,
may also explain these experimental results but are considered unlikely
due to their high formation energy.\citep{Lany_PhysRevLett.98.045501,Chatratin_2019}
In\textsubscript{2}O\textsubscript{3} also possesses a few-nanometers-thick,
two-dimensional surface electron accumulation layer (SEAL), the existence
of which has been demonstrated by a downward band bending at the surface
and the presence of quantized subbands.\citep{king_SEAL-CNL-In2O3_2008,zhang_in2o3_Vo-2013,jovic_10.1002/smll.201903321}
The conductivity of the SEAL is a particularly important property
for applications of In\textsubscript{2}O\textsubscript{3} as gas-sensing
material\citep{rombach_SAB-2016}, facilitates the formation of ohmic
contacts but makes the formation of Schottky contacts challenging.\citep{Michel2019}
Various methods allow the modulation of the SEAL: A reduction of the
SEAL carriers can be achieved by an oxygen plasma treatment of the
surface at elevated temperatures\citep{bierwagen2011PLOX,berthold_plasma_2016,rombach_SAB-2016},
by adsorption of oxidizing reactive species, such as O\textsubscript{2}
and O\textsubscript{3}\citep{Berthold_PSSB_2018} or by intentional
compensating acceptor doping.\citep{Papadogianni_PRB2020}. Conversely,
annealing treatments\citep{bierwagen2011PLOX,berthold_plasma_2016}
or illumination with above-bandgap light\citep{rombach_SAB-2016,Berthold_PSSB_2018}
can enhance the SEAL. Beyond the unintentional conductivity of In\textsubscript{2}O\textsubscript{3},
highly Sn-doped In\textsubscript{2}O\textsubscript{3}\textemdash commonly
known as indium-tin oxide (ITO)\citep{chae2001,tiwari2004,tsai_2016}\textemdash is
commercially used as a transparent contact in optoelectronics, such
as displays, light-emitting diodes, and solar cells.

Ga\textsubscript{2}O\textsubscript{3}, on the other hand, has several
polymorphs,\citep{Roy_JACS.10.1021/ja01123a039} the most stable out
of which is its \textit{\textgreek{b}}-phase with a monoclinic crystal
structure. It has a band gap of approximately 4.8~eV\citep{Matsumoto_1974}
and is, thus, transparent within both the visible and well within
the UV range. Contrary to In\textsubscript{2}O\textsubscript{3},
MBE-grown Ga\textsubscript{2}O\textsubscript{3} is insulating at
room temperature,\citep{Wong_JpnJApplPhys55-2016} even though hydrogen
has been shown to be a shallow donor in Ga\textsubscript{2}O\textsubscript{3}
as well.\citep{Varley_doi:10.1063/1.3499306} As for the nature of
$V_{\mathrm{O}}$, it is widely accepted that they behave as deep
donors in Ga\textsubscript{2}O\textsubscript{3}. Moreover, Ga\textsubscript{2}O\textsubscript{3}
typically exhibits upward band bending\citep{Lovejoy_2012-Ga2O3_Vbb,NAVARROQUEZADA2015368}
and, hence, a surface electron depletion layer, providing challenges
for the formation of ohmic contacts.\citep{Pearton2018}

\begin{figure}
\includegraphics[width=8.6cm]{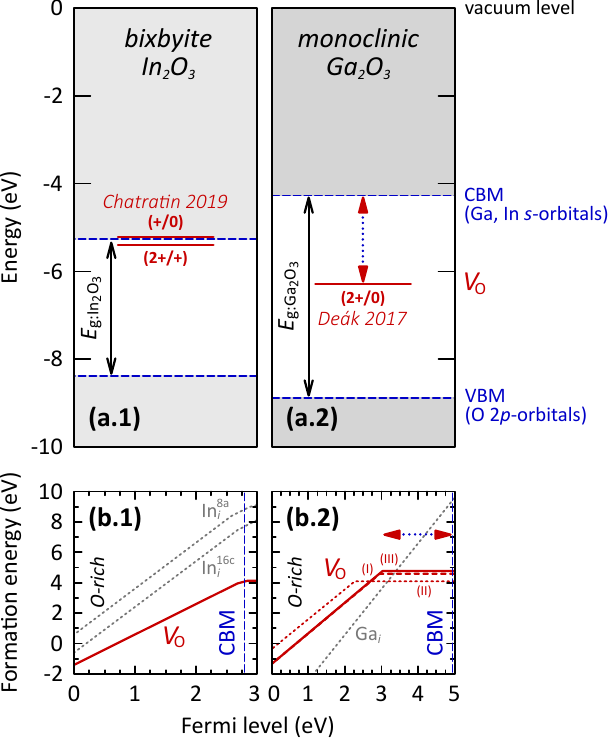}\centering

\caption{(a) Schematic representation of the band edge and relative oxygen
vacancy ($V_{\mathrm{O}}$) levels of cubic In\protect\textsubscript{2}O\protect\textsubscript{3}
and\textit{ \textgreek{b}}-Ga\protect\textsubscript{2}O\protect\textsubscript{3}
based on the calculations of \citet{peelaers2015}. (b) Formation
energies for donorlike point defects under O-rich conditions. Only
the defects with the lowest formation energies and charge state transitions
close to the CBM are depicted. These correspond to $V_{\mathrm{O}}$
(red) and metal interstitials (grey), based on the calculations of
\citet{Chatratin_2019} and \citet{Deak_2017}, respectively. In\textit{
\textgreek{b}}-Ga\protect\textsubscript{2}O\protect\textsubscript{3},
oxygen and, thus, $V_{\mathrm{O}}$ can be found at three different
sites of the unit cell, and the respective formation energies and
(+2/0) transition levels are marked with I, II, and III, as in Ref.~\textcolor{black}{\onlinecite{Chatratin_2019}}.
The dashed blue line corresponds to the CBM at room temperatures.\label{fig:Vo_levels}}
\end{figure}

In theory, a combination of these two compounds can lead to a modulation
of their original properties, such as the position in the band gap\textemdash and
consequently activation energy\textemdash of the $V_{\mathrm{O}}$,
the surface electron density or even the presence of the SEAL altogether.
Figure~\ref{fig:Vo_levels}~(a) graphically illustrates the current
knowledge about the relative band-edge positions of cubic In\textsubscript{2}O\textsubscript{3}
and \textit{\textgreek{b}}-Ga\textsubscript{2}O\textsubscript{3},
along with the charge state transition levels of $V_{\mathrm{O}}$
with respect to the conduction band minimum (CBM). These have been
extrapolated from the formation energy diagrams of \citet{Chatratin_2019}
and \citet{Deak_2017}, depicted schematically in Fig~\ref{fig:Vo_levels}~(b).
An increase of the $V_{\mathrm{O}}$ activation energy would be particularly
interesting, as it could possibly benefit the gas sensitivity in In\textsubscript{2}O\textsubscript{3},
by reducing both the parallel, parasitic contribution of the bulk
conductivity to that of the gas-sensitive surface and the concentration
of adsorption centers. A modulation of the SEAL transport properties
could also benefit this same application. 

Experimentally, a decrease of the Hall electron concentration and
mobility of MOCVD-grown (In\textsubscript{1-\textit{x}}Ga\textsubscript{\textsl{x}})\textsubscript{2}O\textsubscript{3}
films grown on well-matched ZrO\textsubscript{2}:Y (YSZ) substrates
has been earlier demonstrated by \citet{kong_2010} for increasing
$x$ up to 0.9. However, limited information is given on the phase
purity: the films of this study surprisingly remain cubic up to $x=0.5$,
which contradicts current knowledge about the transition from cubic
to monoclinic phase\citep{Wouters_IGO_2020} and may be explained
by a lower incorporation of Ga in the cubic phase than the given total
Ga content. Recent investigations by \citet{Nagata_2020} and \citet{Swallow_2021}
on films grown using pulsed layer deposition (PLD) show the existence
of a SEAL up to approximately $x=0.4$. The latter also demonstrates
the lowering of the charge neutrality level with respect to the conduction
band minimum from In\textsubscript{2}O\textsubscript{3} to InGaO\textsubscript{3}
to Ga\textsubscript{2}O\textsubscript{3} and, hence, the reduction
of the donorlike defects with Ga. However, these studies do not explore
the transport properties and, particularly, the electron mobility
of those films.\textcolor{magenta}{{} }\textcolor{black}{Previously,
In}\textsubscript{\textcolor{black}{2}}\textcolor{black}{O}\textsubscript{\textcolor{black}{3}}\textcolor{black}{{}
(and Ga}\textsubscript{\textcolor{black}{2}}\textcolor{black}{O}\textsubscript{\textcolor{black}{3}}\textcolor{black}{)
films grown by PLD have been shown to exhibit comparably low electron
mobilities,\citep{Borgersen_2020} likely related to high concentrations
of defects.}

Another interesting prospect for (In\textsubscript{1-\textit{x}}Ga\textsubscript{\textsl{x}})\textsubscript{2}O\textsubscript{3}
alloys for applications requiring transparent conductive materials
with a larger band gap than that of ITO can be explored through additional
doping with Sn (IGTO). Recent, application-oriented studies towards
this direction have demonstrated, for instance, polycrystalline or
even amorphous IGTO thin film transistors with high field-effect mobilities.\citep{Kim_IGTO_2021,electronics10111295}\textcolor{red}{{}
}IGTO has been further proposed as an alternative transparent conducting
electrode for light-emitting diodes (LEDs), with the intention to
increase their external quantum efficiency in the near-ultraviolet
spectral range, due to its increased absorption edge, compared to
ITO.\citep{Kim_Kim_2015} %

Recently, we have studied well-defined, high-quality, single-crystalline
\textcolor{black}{(In}\textsubscript{\textcolor{black}{1-}\textit{\textcolor{black}{x}}}\textcolor{black}{Ga}\textsubscript{\textsl{\textcolor{black}{x}}}\textcolor{black}{)}\textsubscript{\textcolor{black}{2}}\textcolor{black}{O}\textsubscript{\textcolor{black}{3}}
layers at the low-\textit{x} end grown by plasma-assisted molecular
beam epitaxy (PA-MBE)\textcolor{black}{{} in terms of surface and film
morphology, crystalline quality, and homogeneity}\textcolor{blue}{\citep{Papadogianni_IGO_structure_2021}}
as well as increasing band gap and optical absorption edge with $x$.\citep{Feldl_2021}

In this work we investigate the effect of Ga on the unintentional
conductivity of the bulk and the SEAL and demonstrate intentional
Sn-doping.

\section{Experimental details}

For the purposes of this study, high quality (111)-oriented single-crystalline
(In\textsubscript{1-\textit{x}}Ga\textsubscript{\textit{x}})\textsubscript{2}O\textsubscript{3}
films, including one (In\textsubscript{0.92}Ga\textsubscript{0.08})\textsubscript{2}O\textsubscript{3}:Sn
film, were synthesized by PA-MBE on quarters of 2-in. insulating YSZ
(111) substrate\textcolor{black}{s, whereas full 2-in. }insulating\textcolor{black}{{}
Al}\textsubscript{\textcolor{black}{2}}\textcolor{black}{O}\textsubscript{\textcolor{black}{3}}\textcolor{black}{{}
(0001) (}\textit{\textcolor{black}{c}}\textcolor{black}{-plane Al}\textsubscript{\textcolor{black}{2}}\textcolor{black}{O}\textsubscript{\textcolor{black}{3}}\textcolor{black}{)
substrates have been employed for the growth of textured samples.
The study focuses on the low-}\textit{\textcolor{black}{x }}\textcolor{black}{($x\leq0.18$)
bixbyite phase end of (In}\textsubscript{\textcolor{black}{1-}\textit{\textcolor{black}{x}}}\textcolor{black}{Ga}\textsubscript{\textit{\textcolor{black}{x}}}\textcolor{black}{)}\textsubscript{\textcolor{black}{2}}\textcolor{black}{O}\textsubscript{\textcolor{black}{3}}\textcolor{black}{,
hence the substrate choice is based on suitability for heteroepitaxy
}of pure In\textsubscript{2}O\textsubscript{3}. After growth, all
samples have been further cleaved into smaller pieces with a size
of approximately 5 \texttimes{} 5 mm\textsuperscript{2}. The total
thickness of the films rang\textcolor{black}{es between 310~nm (SII)
and 700~nm {[}SIII, including the }(In\textsubscript{0.92}Ga\textsubscript{0.08})\textsubscript{2}O\textsubscript{3}:Sn
film\textcolor{black}{{]}. }Detailed information on the growth of
these single-crystalline samples and the determination of the Ga content
therein are reported in our recent work in Ref.~\textcolor{black}{\onlinecite{Papadogianni_IGO_structure_2021}}\textcolor{blue}{.}

\textcolor{black}{The films on }\textit{\textcolor{black}{c}}\textcolor{black}{-Al}\textsubscript{\textcolor{black}{2}}\textcolor{black}{O}\textsubscript{\textcolor{black}{3}}\textcolor{black}{\textemdash not
described in our previous work\textemdash have been grown under similar
conditions upon a pure In}\textsubscript{\textcolor{black}{2}}\textcolor{black}{O}\textsubscript{\textcolor{black}{3}}\textcolor{black}{{}
buffer layer at the interface. }The nucleation of the buffer layer
was conducted under an oxygen plasma flux of 2~standard cubic centimeters
per minute (SCCM), whereas the main part of the film was grown under
a flux of $0.5\,\mathrm{SCCM}$. The power of the oxygen plasma source
was maintained at 300~W throughout the entire growth procedure. The
In-cell temperature was consistently at $850\,\mathrm{{^\circ}C}$
for all samples, corresponding to a beam equivalent pressure (BEP)
of $4\times10^{-7}\,\mathrm{mbar}$. The Ga-containing layers were
grown using varying Ga-cell temperatures from $790\,\mathrm{{^\circ}C}$
up to \textcolor{black}{$840\,\mathrm{{^\circ}C}$ with respective
BEPs ranging between $1.1-3.5\times10^{-8}\,\mathrm{mbar}$}. Under
these conditions the films developed with a growth rate of approximately
$1\,\mathrm{\text{Å/s}}$. The substrate temperature was maintained
at $750\,\mathrm{{^\circ}C}$ throughout both the nucleation and main
film growth\textemdash as measured by a thermocouple between the substrate
and the substrate heater. This is significantly higher than the substrate
temperature used for growth on YSZ, as wetting and overall growth
is \textcolor{black}{more straightforward on Al}\textsubscript{\textcolor{black}{2}}\textcolor{black}{O}\textsubscript{\textcolor{black}{3}}\textcolor{black}{.
The oxygen flow was discontinued after the completion of the growth
from $600\,\mathrm{{^\circ}C}$ downwards and the cool-down was essentially
realized in vacuum. In contrast to the growth procedure on YSZ, an
faster cool-down rate of $0.5\,\mathrm{{^\circ}C/s}$ was employed,
as there is no concern for film delamination in the case of growth
on }\textit{\textcolor{black}{c}}\textcolor{black}{- Al}\textsubscript{\textcolor{black}{2}}\textcolor{black}{O}\textsubscript{\textcolor{black}{3}}\textcolor{black}{. }

Based on the fact that the oxygen chemical potential is higher in
an O-rich environment \citep{Reunchan_2011,Chatratin_2019} and it
decreases with increasing temperature,\citep{Reuter_2001} the free
electron concentrations of the films have been intentionally varied
by annealing treatments in oxygen and vacuum in order to investigate
the effect of Ga on transport. Decrease in the electron concentration
has been achieved through exposure of the films in an oxidizing environment
at high temperature, hereafter referred to as O-annealing. This treatment
was performed within a rapid thermal annealing (RTA) system at a final
temperature of $800\,\mathrm{{^\circ}C}$ at atmospheric pressure
for 60~s, excluding the heating and cooling down processing time.
\textcolor{black}{Conversely, in order to increase the electron concentration
of the films after the initial O-annealing, vacuum annealing treatments
were performed}\textcolor{red}{. }The vacuum annealing was performed
in a magnetron sputter system at a background pressure of approximately
$5\times10^{-7}\,\mathrm{mbar}$ (which increases up to $10^{-3}\,\mathrm{mbar}$upon
annealing) for 5 min at a temperature of approximately $500\,\mathrm{{^\circ}C}$.
The only exception to this was the Sn-doped film, the vacuum-annealing
of which was performed in the growth chamber of the PA-MBE system,
under UHV conditions at background pressure around $5\times10^{-9}\thinspace\mathrm{Torr}$
at a substrate temperature of $650\thinspace\mathrm{{^\circ}C}$ for
a total annealing time of 5~min, excluding the heating and cooling
times (ramping rate of $0.25\,{^\circ}\mathrm{C/s}$ from and to room
temperature, respectively).

Room temperature Hall measurements in the van der Pauw geometry with
non-annealed indium contacts in the corners of the samples provided
the sheet transport properties, which could be translated into average
volume transport properties when combined with the film thickness.

\textcolor{black}{The presence of the SEAL of the films was identified
by a combination of conventional soft x-ray photoelectron spectroscopy
(SXPES: $hv=1.49\,\mathrm{keV}$) using an Al $\mathrm{K\alpha}$
light source and hard x-ray photoelectron spectroscopy (HAXPES: $hv=5.95\,\mathrm{keV}$),
as also documented in our recent works}\textcolor{blue}{.\citep{Papadogianni_IGO_structure_2021,Feldl_2021}}
HAXPES measurements were performed at room temperature at the revolver
undulator beamline at BL15XU of SPring-8.\textcolor{black}{\citep{Ueda_AIPConfProc_2010}}\textcolor{blue}{{}
}A\textcolor{black}{{} detailed description of the experimental setup
of HAXPES at the beamline was described elsewhere.\citep{UEDA2013235}
SXPES measurements were performed with a wide acceptance hemispherical
electron analyzer combined with a monochromatized Al K\textgreek{a}
x-ray source (Thermo Scientific, Sigma Probe). A high-resolution hemispherical
electron analyzer (VG Scienta R4000) was used to detect the photoelectrons.
The total energy resolutions of SXPES and HAXPES were set to 700~meV
and 240~meV, respectively. To determine the absolute binding energy,
the XPES data were calibrated against the Au 4}\textit{\textcolor{black}{f}}\textcolor{black}{}\textsubscript{\textcolor{black}{7/2}}\textcolor{black}{{}
peak (84.0~eV) and the Fermi level of Au. Peak fitting of the XPES
data was carried out using the Voigt function after subtracting the
Shirley-type background.\citep{Shirley_PhysRevB.5.4709} To investigate
the surface band bending behavior, the take-off angle (TOA: $\theta$)
dependent PES was performed. TOA is an angle between the normal vector
to the sample surface and the detector direction. A high angle PES
is surface sensitive, while a low angle it is bulk sensitive. The
corresponding inelastic mean free path (IMFP) of HAXPES and SXPES
for the In 3}\textit{\textcolor{black}{d}}\textcolor{black}{{} core-level
photoemission calculated by the Tanuma\textendash Powell\textendash Penn-2M\citep{Tanuma_2006}
are $\lambda_{\mathrm{HAXPES}}=7.29\,\mathrm{nm}$ and $\lambda_{\mathrm{SXPES}}=2.39\,\mathrm{nm}$,
respectively, meaning that HAXPES probes three times deeper than SXPES.
The probing depth is three times the IMFP,\citep{POWELL19991} therefore,
our HAXPES measurements probe approximately $22\,\mathrm{nm}$ below
the sample surface, which can reduce the effect of the surface Fermi
level pining of In}\textsubscript{\textcolor{black}{2}}\textcolor{black}{O}\textsubscript{\textcolor{black}{3}}\textcolor{black}{.}
To estimate the band bending behaviors of In\textsubscript{2}O\textsubscript{3},
a detailed XPES spectral analysis was performed using COMPRO (Common
Data Processing System) version 11 written by Yoshihara and Yoshikawa,
which simulates the potential energy distribution using the convolution
of calculated peaks at several TOAs.\citep{compro,Yoshihara_2006,Imura_2013}
The simulated spectra were obtained to reproduce the experimental
SXPES and HAXPES spectra at several TOAs by adjusting the band bending
profiles and minimizing the chi squares of the difference between
the simulated and experimental spectra. The band bending direction
was denoted as downward and upward, i.e., towards lower and higher
binding energies, respectively.

\section{Results and discussion}

\begin{figure}
\includegraphics[width=8.6cm]{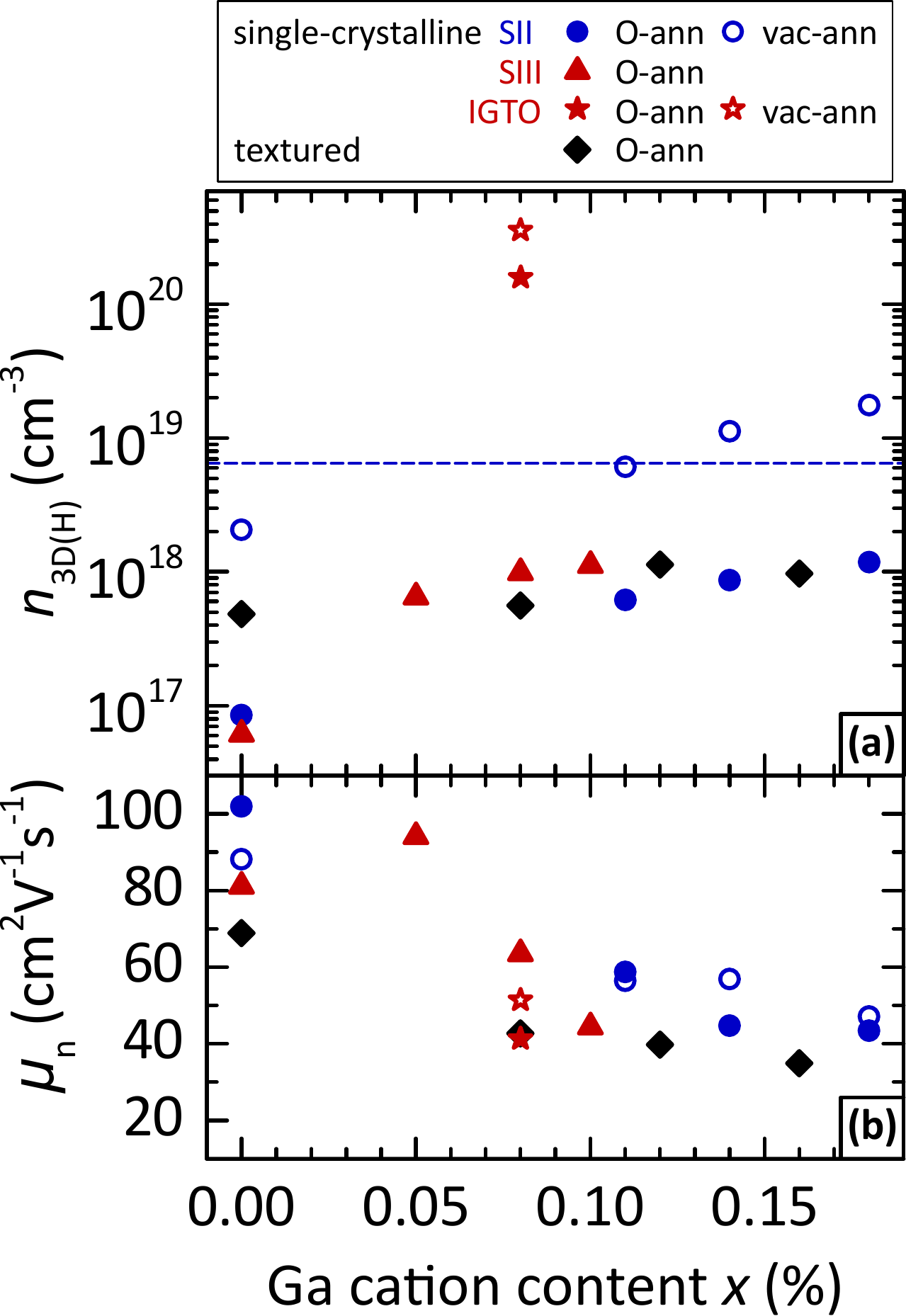}\centering

\caption{Transport properties of single-crystalline samples on YSZ substrates
and textured ones on \textit{c}-Al\protect\textsubscript{2}O\protect\textsubscript{3}:
$n_{\mathrm{3D(\text{\textgreek{H})}}}$ in (a) is the volume electron
concentration of the films extracted by dividing the sheet electron
concentration from Hall effect measurements by the entire film thickness
and $\mu_{\mathrm{n}}$ in (b) is the corresponding Hall electron
mobility. The samples of series SII and SIII have been grown at different
temperatures as described in Ref.~\textcolor{black}{\onlinecite{Papadogianni_IGO_structure_2021}}.
Two different states of the films are investigated: oxygen-annealed
(O-ann), and vacuum-annealed (vac-ann). The dashed blue line in (a)
corresponds to the maximum possible volume electron concentration
due to $V_{\mathrm{O}}$, assuming a charge state transition level
as indicated by Ref.~\textcolor{black}{\onlinecite{Chatratin_2019}}.
The star symbols correspond to an \textcolor{black}{(In}\protect\textsubscript{\textcolor{black}{0.92}}\textcolor{black}{Ga}\protect\textsubscript{\textcolor{black}{0.08}}\textcolor{black}{)}\protect\textsubscript{\textcolor{black}{2}}\textcolor{black}{O}\protect\textsubscript{\textcolor{black}{3}}:Sn
sample (IGTO).\label{fig:transport}}
\end{figure}

Figure~\ref{fig:transport} shows the transport properties of the
single-crystalline and textured films, as determined by four-point
Hall effect measurements in the van der Pauw arrangement.

\subsection{Unintentional bulk doping}

\subsubsection*{Electron concentration}

The electron concentrations, shown in Fig.~\ref{fig:transport}\,(a),
on the order of $\sim10^{17}$\,cm$^{-3}$ and $\sim10^{18}$\,cm$^{-3}$
for the pure In$_{2}$O$_{3}$ films (Ga content $x=0$) annealed
in oxygen and vacuum, respectively, are in good agreement with our
previous findings,\citep{Bierwagen_doi:10.1063/1.4751854} and confirm,
that an oxygen-deficiency provides unintentional shallow donors, e.g.,
oxygen vacancies. With increasing Ga-content, we observe a slight
increase of the electron concentration of the films annealed in oxygen,
which may, in principle, be related to impurities in the Ga-source
that act as shallow donors in the films.

For the single-crystalline films annealed in vacuum, however, the
electron concentration is significantly higher and its increase with
Ga-content is significantly stronger, which cannot be explained by
such impurities and, instead, should be related to point defects due
to an oxygen-poor stoichiometry of the films caused by the annealing.%
{} A plausible explanation for this unanticipated increase in the Hall
electron density $n_{\mathrm{3D(\text{\textgreek{H})}}}$ could be
the formation of donorlike native lattice defects driven by Ga-induced
unit cell distortions.

As discussed, it has been shown that the native lattice defects with
the lowest formation energy in In\textsubscript{2}O\textsubscript{3}
that can act as donors are $V_{\mathrm{O}}$. An O-annealing treatment
can potentially fill in $V_{\mathrm{O}}$ in the films and reduce
$n_{\mathrm{3D(H)}}$. Fig.~\ref{fig:transport}~(a) shows exactly
this effect, if one compares the O-annealed to the vacuum-annealed
state of the films. Since the degree of $V_{\mathrm{O}}$-filling
upon O-annealing is not controllable, it is not necessary that the
O-annealed Ga-containing samples display a decreased $n_{\mathrm{3D(H)}}$
compared to the undoped case. However, assuming the (+/0) transition
level for $V_{\mathrm{O}}$ suggested by \citet{Chatratin_2019} to
correspond to the maximum achievable Fermi level in In\textsubscript{2}O\textsubscript{3}
due to $V_{\mathrm{O}}$ and considering the approximation of the
generalized Einstein relation for degenerate and non-degenerate semiconductors
of \citet{Nilsson}, the resulting electron concentration should never
exceed $6.5\times10^{18}\,\mathrm{cm^{-3}}$ {[}indicated by the horizontal
dashed line in Fig.~\ref{fig:transport}\,(a){]}, for an effective
electron mass of $0.18\,m_{\mathrm{e}}$ as in Ref.~\textcolor{black}{\onlinecite{Feneberg_PhysRevB.93.045203}}.%
{} All vacuum-annealed Ga-containing samples demonstrate equal or higher
$n_{\mathrm{3D(H)}}$ than this upper limit. Moreover, the strongly
increasing $n_{\mathrm{3D(H)}}$ with increasing Ga-content contradicts
the expected lowering of the maximum free electron concentration that
can be provided by $V_{\mathrm{O}}$ due to their expected decreasing
charge transition level with increasing Ga-content.

Unless this charge transition level is significantly higher than theoretically
predicted by \citet{Chatratin_2019}, the hypothesis that the $n_{\mathrm{3D(H)}}$
increase is due to $V_{\mathrm{O}}$ alone does not suffice and other
native donor impurities need to be considered. Those with the next
lowest formation energies are metal interstitials acting as triple-donors
($\mathrm{In}_{i}^{3+}$\citep{Chatratin_2019}, $\mathrm{Ga}_{i}^{3+}$\citep{Deak_2017,Zacherle_2013,Varley_doi:10.1063/1.3499306,Hajnal_1999}).
The neutral charge state transition levels for those seem to be well
within the conduction band and they, thus, serve as a conceivable
explanation for the particularly high electron concentrations observed
here. In addition, the smaller ionic size of $\mathrm{Ga}^{3+}$ than
$\mathrm{In}^{3+}$ makes its incorporation as interstitial more likely,
which would also be consistent with the increasing electron concentration
with increasing Ga-content.%

\subsubsection*{Electron mobility}

\begin{figure}
\includegraphics[width=7.6cm]{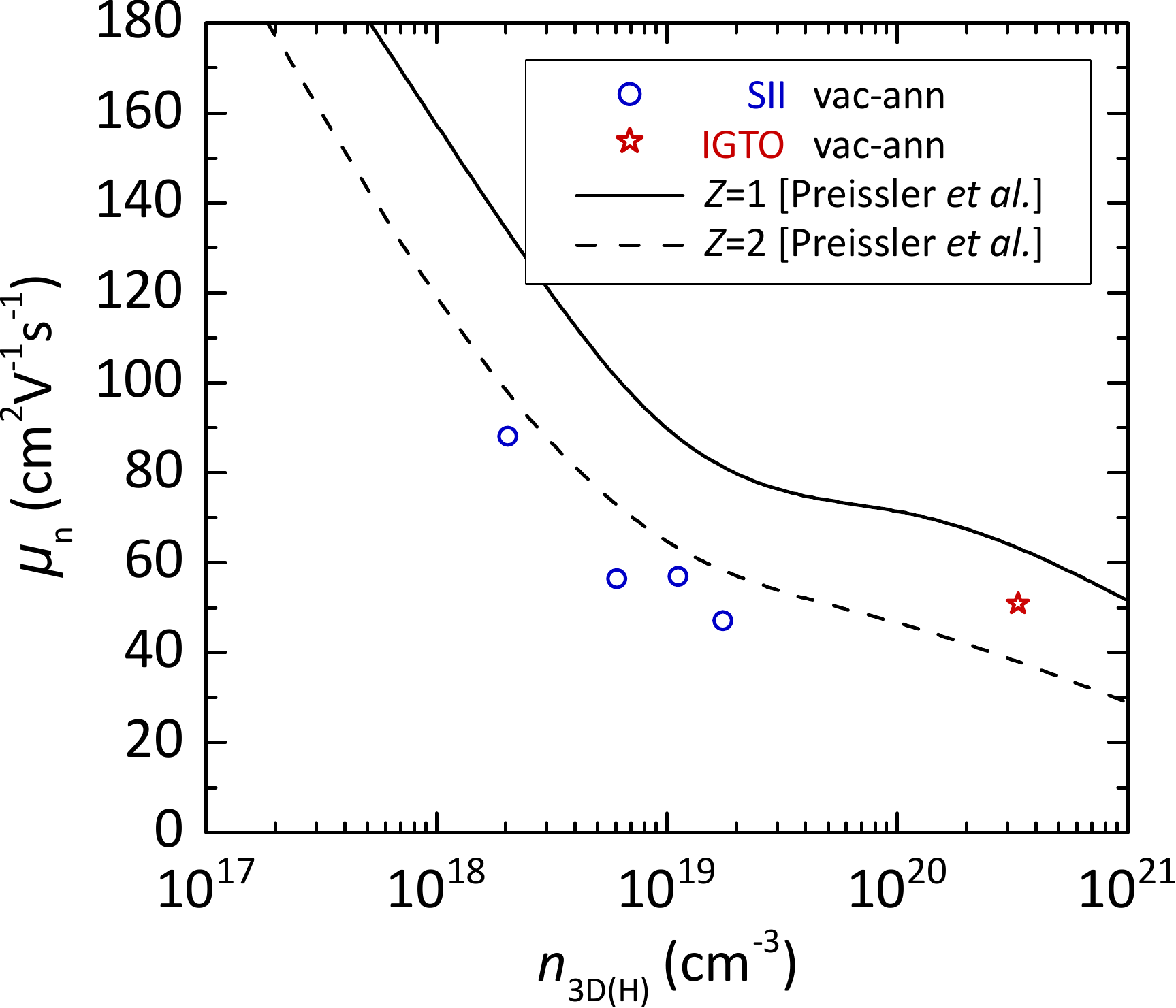}\centering

\caption{Hall electron mobility, $\mu_{\mathrm{n}}$, as a function of the
volume electron concentration from Hall measurements, $n_{\mathrm{3D(\text{\textgreek{H})}}},$of
the single-crystalline vacuum-annealed samples. The solid and dashed
lines correspond to the relation of room temperature Hall mobility
to volume electron concentration modeled by \citet{preisslerPRB}
for singly- ($Z=1$) and doubly-ionized ($Z=2$) donors, respectively.\label{fig:mu_vs_n3D_ioniz}}
\end{figure}

\textcolor{black}{At a first glance, the trend of decreasing mobility
with increasing Ga-content visible in Fig.~\ref{fig:transport}~(b)
is not particularly surprising, as mobility is generally expected
to decrease with incresasing carrier concentration and, thus, ionized
impurity concentration. However, in the case of single-crystalline
Sn-doped In}\textsubscript{\textcolor{black}{2}}\textcolor{black}{O}\textsubscript{\textcolor{black}{3}}\textcolor{black}{{}
films, mobility has been reported to be higher than in the single-crystalline
films studied here for the same range of electron concentrations ($n_{\mathrm{3D(H)}}:\approx10^{18}-\approx10^{19}\,\mathrm{cm^{-3}}$).\citep{preisslerPRB}
This discrepancy would be consistent with the stronger ionized impurity
scattering of doubly-ionized $V_{O}$ in the (In,Ga)$_{2}$O$_{3}$
films compared to the singly ionized donors in the Sn-doped In}\textsubscript{\textcolor{black}{2}}\textcolor{black}{O}\textsubscript{\textcolor{black}{3}}\textcolor{black}{{}
films, theoretically modeled by }\citet{preisslerPRB}\textcolor{black}{.
}This effect is visualized in Fig.~\ref{fig:mu_vs_n3D_ioniz}, with
the IGTO data point lying between the $Z=1$ (Sn\textsuperscript{+})
and $Z=2$ ( $V_{\mathrm{O}}^{2+}$ ). In fact, the good agreement
between the experimental data from the vacuum-annealed films and the
model curve corresponding doubly-ionized donors implies that the origin
of the unexpectedly high electron concentration in the alloy films
cannot be attributed\textemdash for the most part\textemdash to triply-ionized
metal interstitials, as suggested previously. This finding indicates
that the $V_{\mathrm{O}}$ transition level is indeed significantly
higher than theoretically predicted by \citet{Chatratin_2019}. \textcolor{black}{In
addition, distortion in the homogeneity of the crystal potential is
inevitable in alloys and can limit carrier transport by alloy fluctuation
scattering.}

Substrate choice does not seem to significantly affect the results
in terms of electron concentration. However, decreased mobility is
exhibited by the films grown on \textit{c}-Al\textsubscript{2}O\textsubscript{3},
which happens due to scattering at the grain boundaries of the rotational
domains and is not an effect of the Ga incorporation.

\subsection{Intentional bulk doping with Sn}

An \textcolor{black}{(In}\textsubscript{\textcolor{black}{0.92}}\textcolor{black}{Ga}\textsubscript{\textcolor{black}{0.08}}\textcolor{black}{)}\textsubscript{\textcolor{black}{2}}\textcolor{black}{O}\textsubscript{\textcolor{black}{3}}\textcolor{black}{:Sn
film marked as IGTO in Fig.~\ref{fig:transport} has been realized
as a potential advancement of ITO, with the intention of achieving
both comparably high conductivities due to high electron concentrations
and an increased band-gap necessary for higher photon energy applications.
The enlargement of the band-gap of (In}\textsubscript{\textcolor{black}{1-}\textit{\textcolor{black}{x}}}\textcolor{black}{Ga}\textsubscript{\textit{\textcolor{black}{x}}}\textcolor{black}{)}\textsubscript{\textcolor{black}{2}}\textcolor{black}{O}\textsubscript{\textcolor{black}{3}}\textcolor{black}{{}
with and without the addition of Sn in comparison to pure In}\textsubscript{\textcolor{black}{2}}\textcolor{black}{O}\textsubscript{\textcolor{black}{3}}\textcolor{black}{{}
has been confirmed and is discussed in detail in Ref.~\onlinecite{Feldl_2021}.
In this work we investigate its transport properties, which can be
directly compared to the (In}\textsubscript{\textcolor{black}{0.92}}\textcolor{black}{Ga}\textsubscript{\textcolor{black}{0.08}}\textcolor{black}{)}\textsubscript{\textcolor{black}{2}}\textcolor{black}{O}\textsubscript{\textcolor{black}{3}}\textcolor{black}{{}
sample with the same Ga content. As shown in Fig.~\ref{fig:transport}~(a),
incorporation of Sn dramatically increases the }$n_{\mathrm{3D}}$\textcolor{black}{{}
of the film by more than two orders of magnitude}%
. Doping In\textsubscript{2}O\textsubscript{3} and\textemdash hence\textemdash cubic
(In\textsubscript{1-\textit{x}}Ga\textsubscript{\textsl{x}})\textsubscript{2}O\textsubscript{3}
at the low-\textit{x} end\textemdash with an Sn\textsuperscript{4+}
ion should provide one free electron for conduction\citep{Shigesato_1992}.\textcolor{blue}{{}
}The doping effect of Sn in ITO, however, can be largely compensated
upon O-annealing treatments, as the ones performed on the films in
this study, due to the formation of Sn\textendash oxygen-interstitial
clusters, 2Sn\textsubscript{In}·O\textsubscript{\textit{i}}.\citep{Warschkow_2006}
Interstitial oxygen atoms can be removed under reducing conditions,
which breaks down those complexes and frees two donors per O\textsubscript{\textit{i}}
in the process. Consistent with this defect model, we observed an
increase of the electron concentration from $1.5\times10^{20}$ to
$3.3\times10^{20}$~cm$^{-3}$ when comparing the Sn-doped film annealed
in oxygen to that annealed in vacuum. Together with the simultaneous
increase of electron mobility from 38 to 51~cm$^{2}$/Vs {[}shown
in \textcolor{black}{Fig.~\ref{fig:transport}~(b){]}, the film
resistivity decreases from $1.1\times10^{-3}$ to $3.7\times10^{-4}\,\Omega$cm,
well into the range typically observed for the regime of ITO, rendering
IGTO films a suitable candidate for TCO applications.}%

\subsection{SEAL presence in the (In\protect\textsubscript{1-\textit{x}}Ga\protect\textsubscript{\textit{x}})\protect\textsubscript{2}O\protect\textsubscript{3}
films}

\begin{figure}
\includegraphics[width=8cm]{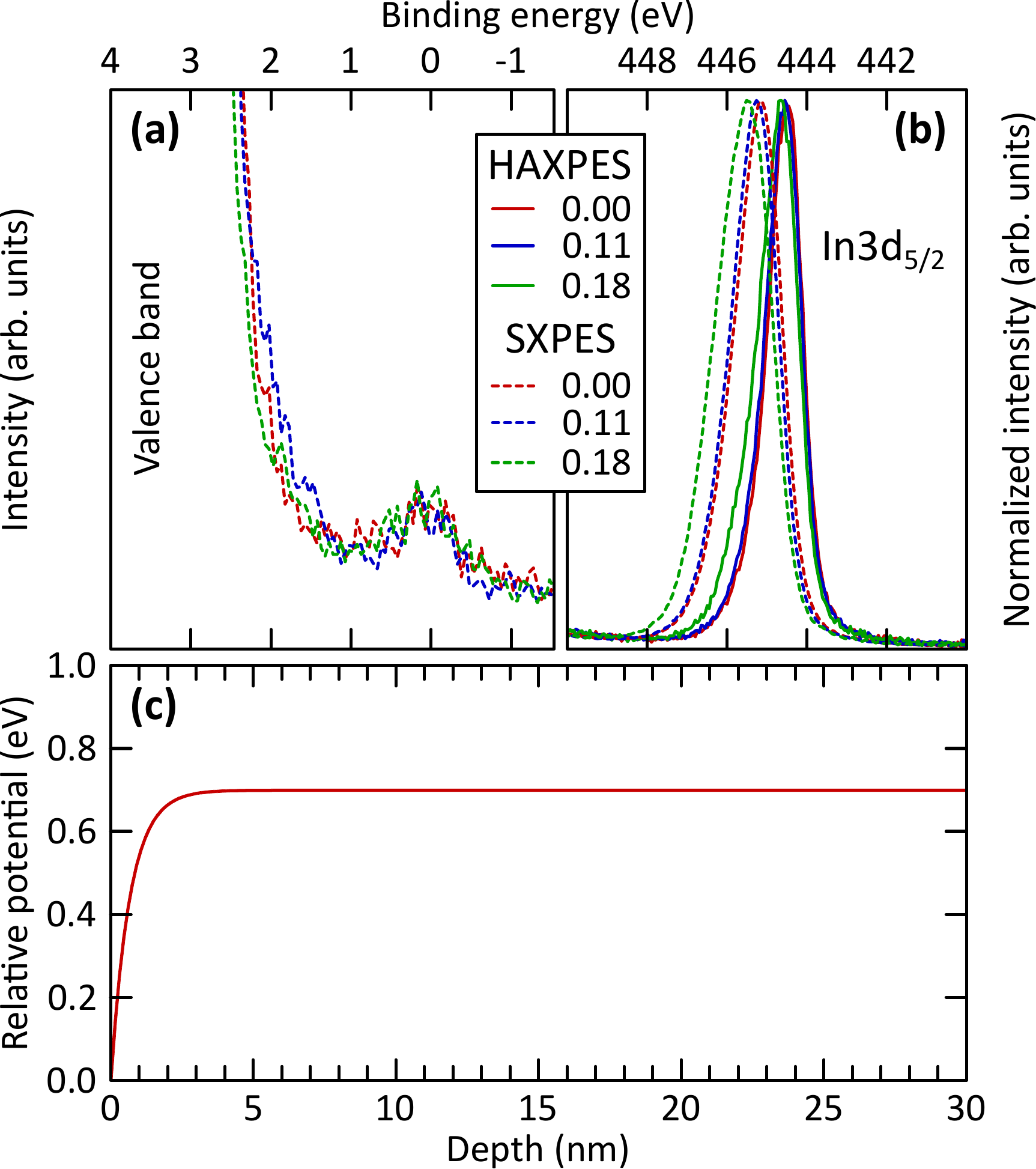}\centering

\caption{(a) Valence band spectrum from SXPES near the Fermi level (0~eV).
(b) In 3\textit{d} core level spectra of the undoped In\protect\textsubscript{2}O\protect\textsubscript{3},
11\,\%, and 18\,\% Ga cation content films obtained by HAXPES (solid
line) at TOA of 5° and SXPES (dashed line) at TOA of 9.7°. (c) Potential
energy distribution as a function depth estimated by COMPRO. The surface
is set at 0~nm.\label{fig:SEAL_SXPES}}
\end{figure}

Magnified SXPES valence band spectra of three representative samples
with Ga content of $x=$0.00, 0.11, and 0.18 (Fig.~\ref{fig:SEAL_SXPES}\,(a))
reveal electronic states at the Fermi level (binding energy of 0~eV),
indicating the presence of the SEAL. Moreover, In 3\textit{d} core
level spectra obtained by SXPES demonstrate a shift towards higher
binding energies than those obtained by HAXPES, as compared in Fig.~\ref{fig:SEAL_SXPES}\,(b).
These results suggest a downward band bending at the surface, a necessary
precondition for the formation of the SEAL. To identify the degree
of the band bending, \textcolor{black}{a detailed spectral analysis
using COMPRO has been performed on the In 3}\textit{\textcolor{black}{d}}\textcolor{black}{{}
spectra obtained by SXPES and HAXPES at various TOA. }Fig.~\ref{fig:SEAL_SXPES}\,(c)
shows potential curve at the surface of the pure In\textsubscript{2}O\textsubscript{3}
film. A drastic potential change has been confirmed within 6~nm from
the surface. All alloy films show similar conduction band states around
the Fermi level and core level shift, which clearly demonstrates the
existence of a SEAL even for samples with Ga cation contents as high
as $x=0.18$.

\section{Conclusion}

Based on current knowledge on the (+/0) charge transition level of
oxygen vacancies in In\textsubscript{2}O\textsubscript{3} and Ga\textsubscript{2}O\textsubscript{3}
and their consequent behavior as shallow and deep donors, respectively,
we have anticipated a decrease in unintentional donor concentration
upon alloying In\textsubscript{2}O\textsubscript{3} with Ga. Nevertheless,
the unintentional electron concentration of the (In\textsubscript{1-\textit{x}}Ga\textsubscript{\textsl{x}})\textsubscript{2}O\textsubscript{3}
alloy films is experimentally shown to increase with Ga\textemdash particularly
for films annealed in vacuum, up to values of $2\times10^{19}$\,cm$^{-3}$.
Hence, we attribute the increased electron concentration to shallow-donorlike
native lattice defects, such as oxygen vacancies or metal\textemdash especially
Ga\textemdash interstitials. According to the reported (+/0) charge
transition level of the oxygen vacancy in the band gap of In\textsubscript{2}O\textsubscript{3},\citep{Chatratin_2019}
this defect cannot be the sole factor contributing to this effect,
as in that case we would not observe electron concentrations higher
than $6.5\times10^{18}\,\mathrm{cm^{-3}}$. Therefore, either the
enhanced carrier concentration is at least partly due to Ga interstitials
or unit cell distortion due to Ga-incorporation induces an increased
incorporation of oxygen vacancies, whose (+/0) charge transition level
is higher than reported in Ref.~\textcolor{black}{\onlinecite{Chatratin_2019}}.
Moreover, the mobility of the alloy films indicates their conductivity
originates from doubly-ionized donors, such as oxygen vacancies, rather
than triply-ionized ones, as the metal interstitials. Hence, the theoretically
calculated charge level transition of the oxygen vacancies needs to
be reconsidered. In addition, the surface electron accumulation layer
native to In\textsubscript{2}O\textsubscript{3}, but not present
in Ga$_{2}$O$_{3}$, does not seem to be affected by the addition
Ga, according to surface-sensitive x-ray photoelectron spectroscopy
measurements. Finally, intentional doping of the alloy films with
Sn has been demonstrated to result in high electron concentrations
up to approximately $3.5\times10^{20}\,\mathrm{cm^{-3}}$ and mobilities
around $50\,\mathrm{cm^{2}/Vs}$. This corresponds to an electrical
resistivity of $\rho=3.5\times10^{-4}\,\mathrm{\Omega cm}$, which
is comparable to that of pure ITO. 

\section*{Acknowledgment}

We would like to thank Jens Herfort for critically reading this manuscript
and Walid Anders for the vacuum annealing processing of the samples.
This study was performed in the framework of GraFOx, a Leibniz-ScienceCampus
partially funded by the Leibniz association. We are grateful to HiSOR,
Hiroshima University, and JAEA/SPring-8 for the development of HAXPES
at BL15XU of SPring-8. The HAXPES measurements were performed under
the approval of the NIMS Synchrotron X-ray Station (Proposal No. 2018A4601,
2018B4600, and 2019B4602).

\bibliographystyle{apsrev4-1}
\bibliography{IGO_transport_literature}

\end{document}